\documentclass[aps,preprint,groupedaddress,showpacs,floatfix]{revtex4}
\usepackage{graphicx}
\begin{document}

\title{Light scalars in semileptonic decays of heavy quarkonia}
\author {
N.N. Achasov$^{\,a}$ \email{achasov@math.nsc.ru} and A.V.
Kiselev$^{\,a,b}$ \email{kiselev@math.nsc.ru}}

\affiliation{
   $^a$Laboratory of Theoretical Physics,
 Sobolev Institute for Mathematics, 630090, Novosibirsk, Russia\\
$^b$Novosibirsk State University, 630090, Novosibirsk, Russia}

\date{\today}

\begin{abstract}

We study the mechanism of production of the light scalar mesons in
the $D_s^+\to\pi^+\pi^-\, e^+\nu$ decays: $D_s^+\to s\bar s\,
e^+\nu\to [\sigma(600)+f_0(980)]\, e^+\nu\to\pi^+\pi^-\, e^+\nu$,
and compare it with  the mechanism of production of the light
pseudoscalar  mesons in the $D_s^+\to (\eta/\eta')\, e^+\nu$
decays: $D_s^+\to s\bar s\, e^+\nu\to (\eta/\eta')\,e^+\nu$. We
show that the $s\bar s\to\sigma(600)$ transition is negligibly
small in comparison with the $s\bar s\to f_0(980)$ one. As for the
the $f_0(980)$ meson, the intensity of the $s\bar s\to f_0(980)$
transition makes near thirty percent from  the intensity of the
$s\bar s\to\eta_s$ ( $\eta_s=s\bar s$ ) transition. So, the
$D_s^+\to\pi^+\pi^-\, e^+\nu$ decay supports the previous
conclusions about a dominant role of  the four-quark components in
the $\sigma(600)$ and $f_0(980)$ mesons.

\end{abstract}

\pacs{13.75.Lb,  11.15.Pg, 11.80.Et,
 12.39.Fe}

\maketitle

At present the  nontrivial nature of the well-established light
scalar resonances $f_0(980)$ and  $a_0(980)$ is  denied by very
few people.  As for the nonet as a whole, even a cursory look at
PDG Review \cite{PDG10} gives an idea of the four-quark structure
of the light scalar meson nonet, $\sigma (600)$, $\kappa (800)$,
$f_0(980)$, and $a_0(980)$, inverted in comparison with the
classical $P$ wave $q\bar q$ tensor meson nonet, $f_2(1270)$,
$a_2(1320)$, $K_2^\ast(1420)$, $\phi_2^\prime (1525)$. Really,
while the scalar nonet  cannot be treated as the $P$ wave $q\bar
q$ nonet in the naive
 quark model, it can be
easy understood as the $q^2\bar q^2$ nonet, where $\sigma$ has no
strange quarks, $\kappa $ has the  s quark, $f_0$ and $a_0$ have
the $s\bar s$ pair. Similar states were found by Jaffe in 1977 in
the MIT bag \cite{jaffe}.

By now it is established also that the mechanisms of the
$a_0(980)$, $f_0(980)$, and $\sigma(600)$ meson production in the
$\phi$ radiative decays
\cite{achasov-89,nna-vvg,achasov-03,nna-avk,nna-avk+,nna-avk++},
in the photon-photon collisions \cite{ads,as}, and in the $\pi\pi$
scattering \cite{nna-avk+,nna-avk++} are the four-quark
transitions and thus indicate to the four-quark structure of the
light scalars \cite{25/9}.

 In addition, the  absence of the $J/\psi$ $\to$ $\gamma f_0(980)$,
$ a_0(980)\rho$, $f_0(980)\omega$ decays in contrast to the
intensive the $J/\psi$ $\to$ $\gamma f_2(1270)$, $ \gamma
f'_2(1525)$, $a_2(1320)\rho$, $f_2(1270)\omega$ decays  argues
against the $P$ wave  $q\bar q$ structure of $a_0(980)$ and
$f_0(980)$ also \cite{achasov-2002}.

\begin{figure}[h]
\begin{center}
\begin{tabular}{ccc}
\includegraphics[width=8cm,height=5cm]{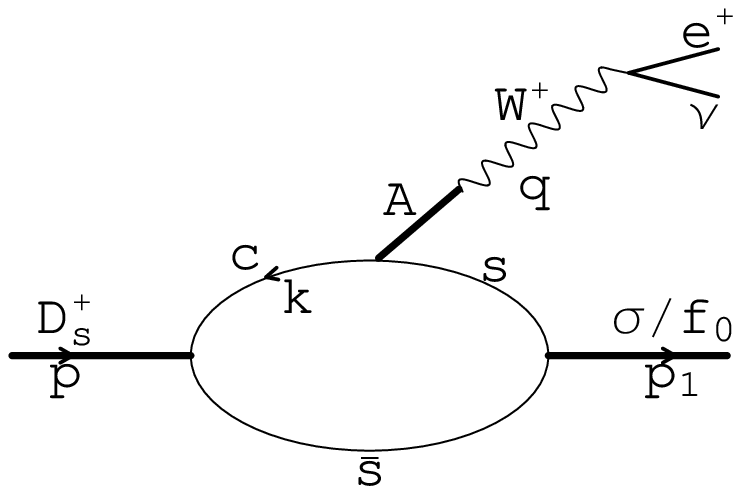}& \includegraphics[width=8cm,height=5cm]{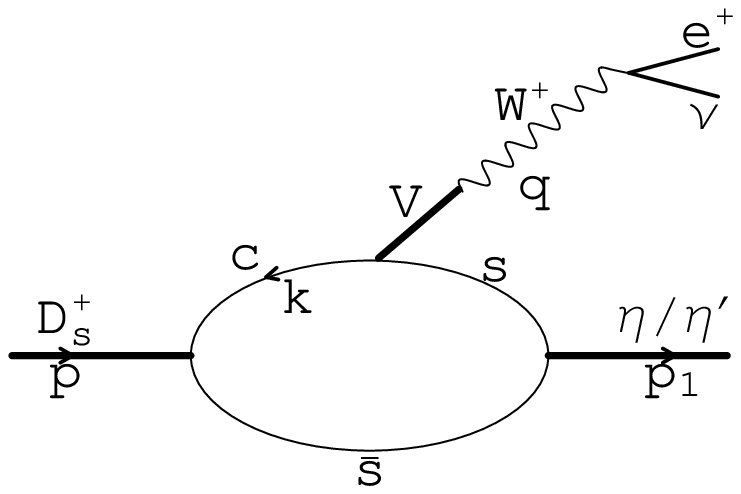}\\ (a)&(b)
\end{tabular}
\end{center}
\caption{Model of the $D^+_s\to\sigma/f_0\, e^+\nu$ and $D^+_s\to
( \eta/\eta')\, e^+\nu$ decays} \label{fig1}
\end{figure}

It is time to explore the light scalar mesons in the decays of of
heavy quarkonia   \cite{cleo,shechter,harada}. The semileptonic
decays are of prime interest because they have the clear
mechanisms, see, for example, Fig. \ref{fig1}.

 As Fig. \ref{fig1}
suggests, the $D_s^+\to s\bar s\, e^+\nu\to
[\sigma(600)+f_0(980)]\, e^+\nu\to\pi^+\pi^-\, e^+\nu$ decay is
the perfect probe of the $s\bar s$ component in the $\sigma(600)$
and $f_0(980)$ states  \cite{cleo,shechter}.

Below we study the mechanism of production of the light scalar
mesons in the $D_s^+\to\pi^+\pi^-\, e^+\nu$ decays: $D_s^+\to
s\bar s\, e^+\nu\to [\sigma(600)+f_0(980)]\, e^+\nu\to\pi^+\pi^-\,
e^+\nu$, and compare it with  the mechanism of production of the
light pseudoscalar  mesons in the $D_s^+\to (\eta/\eta')\, e^+\nu$
decays: $D_s^+\to s\bar s\, e^+\nu\to (\eta/\eta')\,e^+\nu$, in a
model of the Nambu-Jona-Lasinio type \cite{mkv}.

The  amplitudes of the $D_s^+\to P (\mbox{pseudoscalar)}\, e^+\nu$
and $D_s^+\to S (\mbox{scalar)}\, e^+\nu$ decays have the form
\begin{eqnarray}
& M[D_s^+(p)\to P(p_1) W^+(q)\to P(p_1)\,
e^+\nu]=\frac{G_F}{\sqrt{2}}V_{cs}V_\alpha L^\alpha\,, \nonumber\\
 & M[D_s^+(p)\to S(p_1)W^+(q)\to S(p_1)\, e^+\nu]=\frac{G_F}{\sqrt{2}}V_{cs}A_\alpha
 L^\alpha\,,
 \label{amplitudes}
\end{eqnarray}
where $G_F$ is the Fermi constant, $V_{cs}$ is the
Cabibbo-Kobayashi-Maskava matrix element,
\begin{eqnarray}
& V_\alpha = f^P_+(q^2)(p+p_1)_\alpha +
f^P_-(q^2)(p-p_1)_\alpha\,, \nonumber
\\ & A_\alpha = f^S_+(q^2)(p+p_1)_\alpha + f^S_-(q^2)(p-p_1)_\alpha\,,\nonumber \\
 & L_\alpha =\bar{\nu}\gamma_\alpha(1+\gamma_5)e\,,\ \ \ \ \ \ \
 q=(p-p_1)\,.
 \label{VAL}
\end{eqnarray}

The influence of the $f^P_-(q^2)$ and  $f^S_-(q^2)$ form factors
are negligible because of the small mass of the positron.

 The
decay rates in the stable $P$ and $S$ states are
\begin{eqnarray}
&\frac{d\Gamma (D_s^+\to
P\,e^+\nu)}{dq^2}=\frac{G^2_F|V_{cs}|^2}{24\pi^3}p^3_1(q^2)|f^P_+(q^2)|^2,\
\ \frac{d\Gamma(D_s^+\to
S\,e^+\nu)}{dq^2}=\frac{G^2_F|V_{cs}|^2}{24\pi^3}p^3_1(q^2)|f^S_+(q^2)|^2,\nonumber\\[9pt]
& p_1(q^2)=
\frac{\sqrt{m^4_{D^+_s}-2m^2_{D^+_s}(q^2+m^2_P)+(q^2-m^2_P)^2
}}{2m_{D^+_s}}\,,\ \mbox{or}\ \ p_1(q^2)=
\frac{\sqrt{m^4_{D^+_s}-2m^2_{D^+_s}(q^2+m^2_S)+(q^2-m^2_S)^2
}}{2m_{D^+_s}}\,.
 \label{dGammadq2}
\end{eqnarray}

For the $f^P_+(q^2)$ and  $f^S_+(q^2)$ form factors we use the
vector dominance model
\begin{equation}
f^P_+(q^2)=f^P_+(0)\frac{m^2_V}{m^2_V-q^2}=f^P_+(0)f_V(q^2)\,,\ \
\ \ \ \ \ \ \ f^S_+(q^2)=
f^S_+(0)\frac{m^2_A}{m^2_A-q^2}=f^S_+(0)f_A(q^2)\,,
 \label{VDM}
\end{equation}
where $V=D^*_s(2112)^\pm$, $A=D_{s1}(2460)^\pm$, \cite{PDG10}.

Following Fig. \ref{fig1} we write   $f_+^P(0)$ and $f_+^S(0)$ in
the form
\begin{equation}
f_+^P(0)= g_{D_s^+c\bar s}F_Pg_{s\bar sP}\,,\ \ \ \ f_+^S(0)=
g_{D_s^+c\bar s}F_Sg_{s\bar sS}\,,
 \label{definitions}
\end{equation}
where $g_{D_s^+c\bar s}$ is the $D_s^+\to c\bar s$ coupling
constant, $g_{s\bar sP}$ and $g_{s\bar sS}$ are the $s\bar s\to P$
and $s\bar s\to S$ coupling constants.

We know the structure of $\eta$ and $\eta'$
\begin{equation}
 \eta= \eta_q\,\cos\phi - \eta_s\,\sin\phi\,,\ \ \ \ \ \eta'=
\eta_q\,\sin\phi+ \eta_s\,\cos\phi\,,
 \label{etaetaprime}
\end{equation}
where $\eta_q=(u\bar u + d\bar d)/\sqrt{2}$ and $\eta_s =s\bar s$.
The angle $\phi=\theta_i+\theta_P$, where $\theta_i$ is the ideal
mixing angle with $\cos\theta_i=\sqrt{1/3}$ and
$\sin\theta_i=\sqrt{2/3}$, i.e., $\theta_i=54.7^\circ$, and
$\theta_P$ is the angle between the flavor-singlet state $\eta_1$
and the flavor-octet state $\eta_8$.

So,
\begin{equation}
g_{s\bar s\eta}=-  g_{s\bar s\eta_s}\sin\phi\,,\ \ \ \ g_{s\bar
s\eta'}=  g_{s\bar s\eta_s}\cos\phi\,.
 \label{sbarsetaeta'}
\end{equation}

The Particle Data Group \cite{PDG10} gives the $\theta_P$ band
$-20^\circ \lesssim \theta_P\lesssim - 10^\circ$ that gives us the
opportunity to extract information about the $s\bar s\to\eta_s$
coupling constant, $g_{s\bar s\eta_s}$, from  experiment and to
compare with the $s\bar s\to f_0$ coupling constant, $g_{s\bar s
f_0}$, extracted
 from experiment also. We consider
 the next set of $\theta_P$.
 \begin{eqnarray}
& \theta_P=-11^\circ\,:\ \ \ \ \eta=0.72\eta_0-0.69\eta_s\,,\ \ \
\ \eta'=0.69\eta_0+0.72\eta_s\nonumber\\
 & \theta_P=-14^\circ\,:\
\ \ \ \eta=0.76\eta_0-0.65\eta_s\,,\ \ \ \
\eta'=0.65\eta_0+0.76\eta_s\nonumber\\
 & \hspace*{-17pt}
\theta_P=-18^\circ\,:\ \ \ \ \eta=0.8\eta_0-0.6\eta_s\,,\ \ \ \
\eta'=0.6\eta_0+0.8\eta_s\,.
 \label{angles}
 \end{eqnarray}

 The amplitude of the  the $D_s^+\to s\bar s\, e^+\nu\to
[\sigma(600)+f_0(980)]\, e^+\nu\to\pi^+\pi^-\, e^+\nu$ decay is
\begin{eqnarray}
& & M( D_s^+\to s\bar s\, e^+\nu\, \to\pi^+\pi^-\, e^+\nu
)=\frac{G_F}{\sqrt{2}}V_{cs}\,
 L^\alpha\,(p+p_1)_\alpha\,g_{D_s^+c\bar
s}\,f_A(q^2)\nonumber\\[9pt]
 & &\times e^{i\delta_B^{\pi\pi}}\frac{1}{\Delta(m)}\,\Bigl(F_\sigma g_{s\bar s\sigma}D_{f_0}(m)g_{\sigma\pi^+\pi^-}+F_\sigma g_{s\bar s\sigma}
\Pi_{\sigma f_0}(m)g_{f_0\pi^+\pi^-}\nonumber\\[9pt] & & \mbox{} +
F_{f_0} g_{s\bar s f_0}
\Pi_{f_0\sigma}(m)g_{\sigma\pi^+\pi^-}+F_{f_0} g_{s\bar s f_0}
D_\sigma(m)g_{f_0 \pi^+\pi^-}\Bigr )\,,
 \label{Ds+decayamplitude}
\end{eqnarray}
where $m$ is the invariant mass of the $\pi\pi$ system,
$\Delta(m)= D_{f_0}(m)D_\sigma(m)-\Pi_{f_0\sigma}(m)\Pi_{\sigma
f_0}(m)$, $D_\sigma(m)$ and $D_{f_0}(m)$ are the inverted
propagators of the $\sigma$ and $f_0$ mesons, $\Pi_{\sigma
f_0}(m)=\Pi_{f_0\sigma}(m)$ is the off-diagonal element of the
polarization operator, which mixes the $\sigma$ and $f_0$ mesons.
All the details can be found in  Refs.
\cite{nna-avk+,nna-avk++,as}.

 The double differential rate of the $D_s^+\to s\bar s\, e^+\nu\to [\sigma(600)+f_0(980)]\,
e^+\nu\to\pi^+\pi^-\, e^+\nu$ decay is
\begin{eqnarray}
& & \frac{d^2\Gamma(D_s^+\to\pi^+\pi^-\,
e^+\nu)}{dq^2dm}=\frac{G^2_F\,|V_{cs}|^2}{24\,\pi^3}\,g_{D_s^+c\bar
s}^2\,|f_A(q^2)|^2\,p^3_1(q^2,m)\nonumber\\[9pt] & &
\times\,\frac{1}{8\pi^2}\,m\,\rho_{\pi\pi}(m)\,\Bigl|\frac{1}{\Delta(m)}\Bigr|^2\,
\Bigl|F_\sigma g_{s\bar s\sigma}
D_{f_0}(m)g_{\sigma\pi^+\pi^-}+F_\sigma g_{s\bar s\sigma}
\Pi_{\sigma f_0}(m)g_{f_0\pi^+\pi^-} \nonumber\\[9pt]
 & & \mbox{}
  + F_{f_0} g_{s\bar s f_0}
\Pi_{f_0\sigma}(m)g_{\sigma\pi^+\pi^-}+F_{f_0} g_{s\bar s f_0}
D_\sigma(m)g_{f_0 \pi^+\pi^-}\Bigr|^2\,,
 \label{d2Gammmadq^2dm}
\end{eqnarray}
where $\rho_{\pi\pi}(m)=\sqrt{1-4m_\pi^2/m^2}$.

\begin{figure}[h]
\begin{center}
\includegraphics[width=14cm,height=10cm]{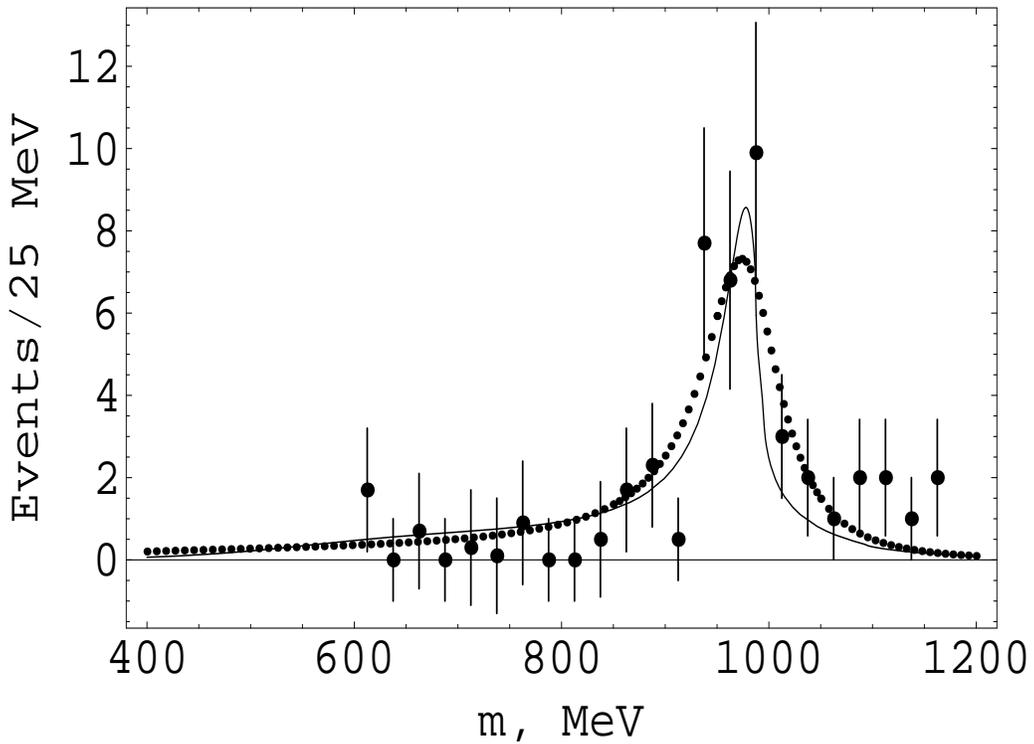}
\end{center}
 \caption{ The CLEO data \cite{cleo} on the
invariant $\pi^+\pi^-$ mass ($m$) distribution for
$D^+_s\to\pi^+\pi^-e^+\nu$ decay with the subtracted backgrounds,
which are calculated in Ref. \cite{cleo}. The dotted line is Fit
from Ref. \cite{cleo}, Fig. 9, corresponding to $BR(D^+_s\to
f_0(980)\,e^+\nu)\,BR(f_0(980)\to\pi+\pi^-) =(0.20\pm 0.03\pm
0.01)$.  Our theoretical curve is the solid line. } \label{fig2}
\end{figure}

\begin{figure}[h]
\begin{center}
\includegraphics[width=12cm,height=10cm]{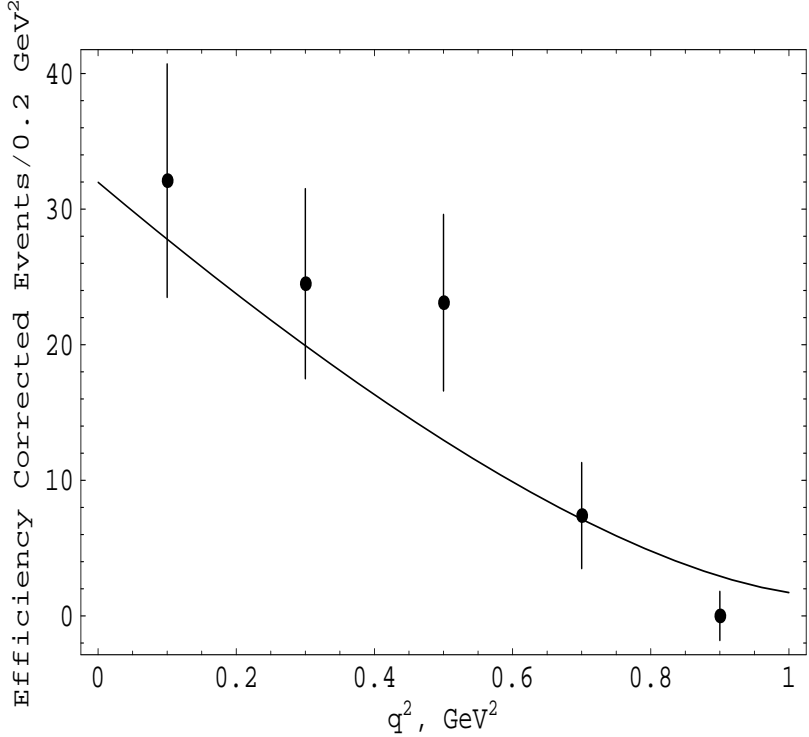}
\end{center}
\caption{ The $q^2$  distribution for $BR(D^+_s\to
f_0(980)\,e^+\nu)$. The axial-vector dominance model, see Eq.
(\ref{VDM}), describes the CLEO data \cite{cleo} quite
satisfactorily. } \label{fig3}
\end{figure}

 When $\Pi_{\sigma f_0}(m)=\Pi_{f_0\sigma}(m)=0$ and
 $g_{s\bar s\sigma}=0$
 \begin{equation}
\frac{d^2\Gamma(D_s^+\to\pi^+\pi^-\,
e^+\nu)}{dq^2dm}=\frac{G^2_F\,|V_{cs}|^2}{24\,\pi^3}\,g_{D_s^+c\bar
s}^2\,|f_A(q^2)|^2\,p^3_1(q^2,m)\,\frac{2}{\pi}\,\frac{m^2\,\Gamma(f_0\to\pi^+\pi^-\,m)}{|D_{f_0}(m)|^2}\,.
  \label{singlef0}
\end{equation}
When fitting the CLEO \cite{cleo}, we use the parameters of the
resonances obtained in Ref. \cite{nna-avk++} in the analysis of
the $\pi\pi$ scattering and the $\phi\to\gamma
(\sigma+f_0)\to\pi^0\pi^0$ decay. So the 44 events in Fig.
\ref{fig2} determine only one parameter
$f_+^{\sigma}(0)/f_+^{f_0}(0)$. In this case the Adler self
consistency condition (the Adler zero at $m^2$ near $(m_\pi^2)/2$)
determines $f_+^\sigma(0)/f_+^{f_0}(0)=(F_\sigma g_{s\bar
s\sigma})/(F_{f_0} g_{s\bar s f_0})$=0.039, 0.014, 0.055, 0.058,
0.032, 0.055 for six fits from Ref. \cite{nna-avk++}.  So the
intensity of the $\sigma(600)$ production is much less than the
intensity of the $f_0(980)$ production
($(f_+^\sigma(0)/f_+^{f_0}(0))^2 \leq 0.003$). That is we find the
direct evidence of decoupling of $\sigma(600)$ with the $s\bar s$
pair. {\bf As far as we know, this is truly a new result}, which
agrees well with the decoupling of $\sigma(600)$ with the $K\bar
K$ states, obtained in Ref. \cite{nna-avk++} $g_{\sigma
K^+K^-}^2/g_{\sigma\pi^+\pi^-}^2$ =0.04, 0.001, 0.01, 0.01, 0.003,
0.025 for six fits. The decoupling of $\sigma(600)$ with the
$K\bar K$ states means also the decoupling of $\sigma(600)$ with
$\sigma_q=(u\bar u + d\bar d)/\sqrt{2}$ because $\sigma_q$ results
in $g_{\sigma K^+K^-}^2/g_{\sigma \pi^+\pi^-}^2=1/4$. Results of
our analysis of the CLEO \cite{cleo} data are shown in the Table
and on Figs. \ref{fig2} and  \ref{fig3}.  The parameters of the
$\sigma(600)$ and $f_0(980)$ mesons are taken from Fit 1 of Ref.
\cite{nna-avk++} which describes the spectrum on Fig. \ref{fig2}
better than others ($(F_\sigma g_{s\bar s\sigma})/(F_{f_0}
g_{s\bar s f_0})=0.039$, $g_{\sigma
K^+K^-}^2/g_{\sigma\pi^+\pi^-}^2 =0.04$).
 So,the CLEO experiment gives new support in favour of the
four-quark ($ud\bar u\bar d$) structure of the $\sigma(600)$
meson.

 In the chirally symmetric model of the
Nambu-Jona-Lasinio type the coupling constants of the pseudoscalar
and scalar partners with quarks are equal to each other, i.e.,
 $g_{s\bar s\eta_s}=g_{s\bar s f_{0s}}$, where $f_{0s}=s\bar s$.
 In approximation when the mass of the strange quark much less the mass of the charmed
quark ($m_s/m_c\ll 1$)  $F_{f_0}=F_{\eta'}$ \cite{ms/mc} and we
find from the Table (see the last line) that $g^2_{s\bar s
f_0}/g^2_{s\bar s\eta_s}\approx 0.3$. So, the $f_{0s}=s\bar s$
part in the $f_0(980)$ wave function is near thirty percent.
Taking into account the suppression  of the $f_0(980)$ meson
coupling with the $\pi\pi$ system, $g_{f_0 \pi^+\pi^-}^2/g_{f_0
K^+K^-}^2=0.154$, see Fit 1 in the Table I of Ref.
\cite{nna-avk++}, one can conclude that the $f_{0q}= (u\bar u +
d\bar d)/\sqrt{2}$ part  in the $f_0(980)$ wave function is
suppressed also. So, the CLEO experiment gives strong support in
favour of the four-quark ($sd\bar s\bar d$) structure of the
$f_0(980)$ meson, too.
\newpage
 \noindent Table. Results of the analysis of the CLEO
\cite{cleo} data. All quantities are defined in the text.

\begin{center}
\begin{tabular}{|c|c|c|c|}\hline
\multicolumn{4}{|c|}{ $Br(D_s^+\to f_0 e^+\nu\to\pi^+\pi^-e^+\nu)=
0.17\%$}
\\ \hline

$(F_\sigma g_{s\bar s\sigma})/(F_{f_0} g_{s\bar s f_0})$&
$(F^2_{f_0}g^2_{s\bar s f_0}) /(F^2_{\eta}g^2_{s\bar s\eta})$ & $(
F^2_{f_0}g^2_{s\bar s f_0}) /(F^2_{\eta'}g^2_{s\bar s\eta'})$ &
$(F^2_{\eta}g^2_{s\bar s\eta}) /(F^2_{\eta'}g^2_{s\bar s\eta'})$\\
\hline

 $0.039$& $0.67$ &
$0.49$ & $0.73$\\ \hline

\multicolumn{4}{|c|}{The $\eta-\eta'$ mixing }\\ \hline

$\theta_P$ & $ -11^\circ$ & $-14^\circ$ & $-18^\circ$  \\ \hline

 $( F^2_{f_0}g^2_{s\bar s f_0}) /(F^2_{\eta}g^2_{s\bar s\eta_s})$ &$ 0.32$ & $ 0.29$ & $ 0.24$

\\ \hline

 $(F^2_{f_0}g^2_{s\bar s
f_0})/(F_{\eta'}^2g^2_{s\bar s\eta_s})$ & $0.27$ & $0.28$& $0.31$
\\ \hline
\end{tabular}
\end{center}
\vspace*{9pt}

\newpage
Certainly, there is an extreme need in experiment on the
$D_s^+\to\pi^+\pi^-\, e^+\nu$ decay with high statistics.

 Of great interest is the experimental search for the
decays $D^0\to d\bar u\,e^+\nu\to a^-_0(980)\,
e^+\nu\to\pi^-\eta\, e^+\nu$ and $D^+\to d\bar d\,e^+\nu\to
a^0_0(980)\, e^+\nu\to\pi^0\eta\, e^+\nu$ (or the charge conjugate
ones),
 which will give the information about the
$a^-_q=d\bar u$ (or $a^+_q=u\bar d$)  and $a^0_q=(u\bar u - d\bar
d)/\sqrt{2}$ components in the $a^-_0(980)$ and $a^0_0$ wave
functions respectively.

No less interesting is also search for the decays $D^+\to d\bar
d\,e^+\nu\to [\sigma(600)+f_0(980)]\, e^+\nu\to\pi^+\pi^-\,
e^+\nu$ (or the charge conjugate ones), which will give the
information about the $\sigma_q = (u\bar u + d\bar d)/\sqrt{2} $
and $f_{0q}= (u\bar u + d\bar d)/\sqrt{2}$ components in the
$\sigma(600)$ and $f_0(980)$ wave functions respectively.\\[9pt]
 This work
was supported in part by RFBR, Grant No 10-02-00016, and
Interdisciplinary project No 102 of Siberian division of RAS.


\begin{thebibliography}{99}
\bibitem{25/9} In particular, the was
  shown that  the ideal  $q\bar q$ model prediction
  $g^2_{f_0(980)\gamma\gamma}:g^2_{a^0_0(980)\gamma\gamma}=25:9$ is excluded by experiment in contrast to the simalar
  prediction for the tensor states $f_2(1270)$ and $a_2(1320)$.
 We mean here the   $f_0(980)=(u\bar u+d\bar
d)\sqrt{2}$ and  $a^0_0(980)=(u\bar u-d\bar d)\sqrt{2}$ case for
equality of the  masses: $m_{f_0}=m_{a_0^0}$\,.
\bibitem{ms/mc} The study beyond this approximation we hope to carry out
subsequently.
\end{thebibliography}
\end{document}